\documentclass[10pt,twocolumn]{article}

\usepackage{stfloats} 

\usepackage[utf8]{inputenc}
\usepackage[letterpaper, margin=0.7in]{geometry} 
\usepackage{newtxtext,newtxmath}                
\usepackage{graphicx}
\usepackage{amsmath}
\usepackage{booktabs}                           
\usepackage{lipsum}                             

\usepackage{abstract}

\title{\textbf{\Large Long-Term \(10^{-18}\)-Level Flexible Stability Transfer with a Hybrid-Locked Kerr Microcomb}}
\author{\large Andrei Diakonov$^{1}$, Konstantin Khrizman$^{1}$ and Liron Stern$^1$ \\
\small\textit{$^1$Institute of Applied Physics, The Hebrew University of Jerusalem, Israel}
}
\date{}

\begin{document}

\twocolumn[
  \maketitle
  \begin{abstract}
  \noindent 

Optical atomic clocks are moving toward deployable systems, with commercial instruments entering the $10^{-17}$ fractional-frequency-instability regime. Transferring this performance to accessible microwave and radio-frequency signals requires compact optical frequency combs that preserve long-term stability during frequency division. Integrated Kerr microcombs have opened a new era in chip-scale precision metrology, enabling optical-frequency division and low-noise microwave generation. Yet, whether they can robustly preserve the stability required by state-of-the-art optical lattice and single-ion clocks over hour-long timescales remains an open question. Here we demonstrate that a hybrid passive-active stabilization of a Kerr microcomb, combining injection-locking of one tooth with active stabilization of a second, freely selected tooth through pump-frequency actuation, can indeed satisfy the metrological requirements at the highest level. The separated control channels effectively orthogonalize stabilization of the two microcomb frequency degrees of freedom while leaving the pump frequency unconstrained. A continuous five-hour measurement, without drift removal, yields a residual relative instability of $1.4\times10^{-16}$ at 1 s and $2\times10^{-18}$ at 4000 s on an out-of-loop tooth. We further show that the residual instability remains unchanged as the injection power is varied over 14 dB, demonstrating robust operation across a broad locking range. These results bring integrated microcomb stabilization into the residual-noise regime required to preserve the long-term performance of emerging optical lattice and single-ion clock systems.

  \vspace{0.5cm}
  \end{abstract}
]

\section*{Main}
Exquisite frequency and timing references, spanning microwave to optical atomic clocks, underpin modern science, navigation, and communications. Increasingly accurate clocks are required for applications including navigation, geodesy \cite{mcgrew2018atomic}, natural-disaster detection \cite{blewitt2009gps}, and fundamental physics, where clocks can probe potential variations in physical constants \cite{safronova2019search}. Microwave atomic clocks generally rely on hyperfine transitions in alkali atoms such as cesium or rubidium, operating at gigahertz frequencies. The use of transitions at progressively higher frequencies enables finer phase discrimination and improved fractional-frequency stability. This principle has been exploited with spectacular success in optical atomic clocks \cite{ludlow2015optical}, where transitions in the hundreds of terahertz have enabled fractional-frequency instabilities at the $10^{-18}$ level \cite{yang2025clock}. Optical frequency combs bridge these otherwise electronically inaccessible optical oscillations to the radio-frequency domain through a series of equally spaced, mutually coherent lines. They act essentially as an optical divider.

\begin{figure*}[h!]
    \centering
    \includegraphics[width=0.95\linewidth]{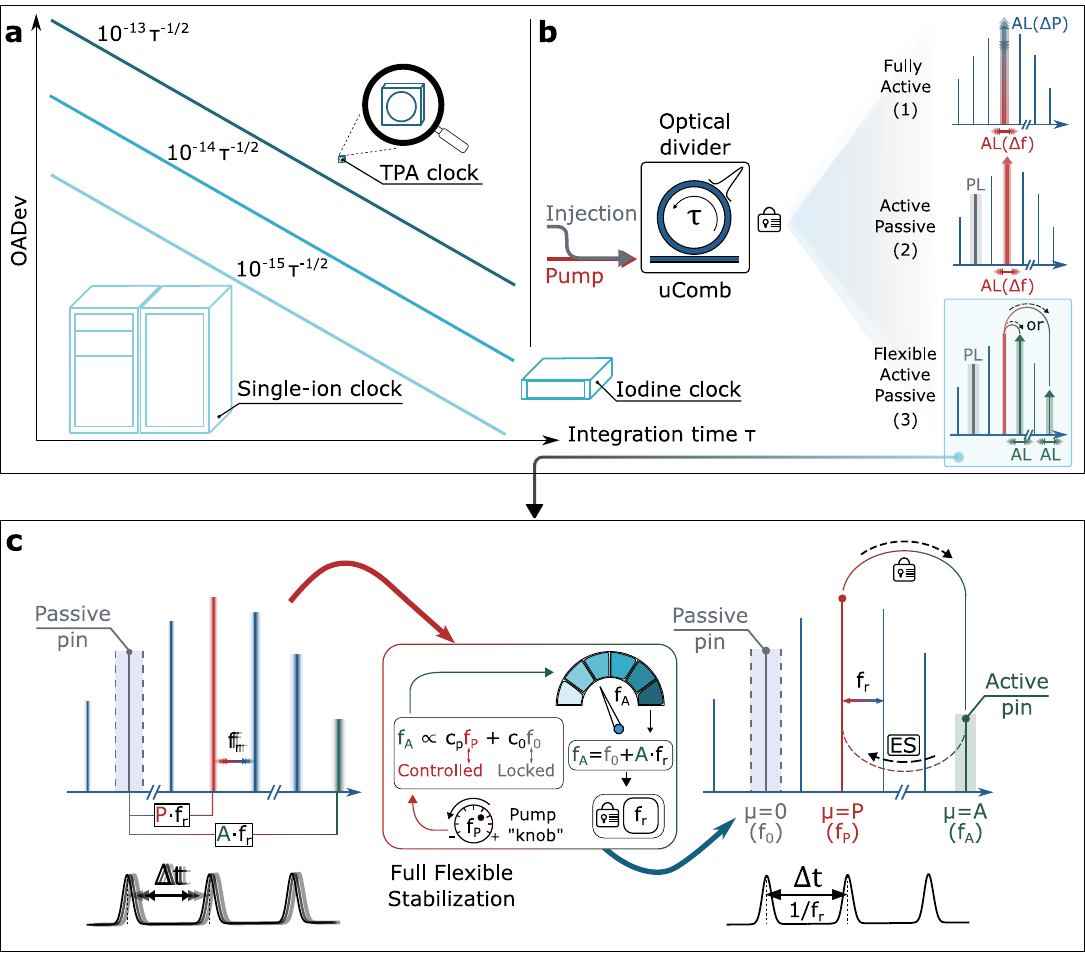}
    \caption{\textbf{Unconstrained hybrid microcomb stabilization for high-end optical atomic clock stability transfer.} a) Comparison of the optical atomic clock architectures stability: single-ion clock, iodine clock and two-photon absorption (TPA) implementations can operate at $10^{-15}, 10^{-14}$ and $10^{-13}$ levels correspondingly, as depicted schematically by their Allan Deviation (ADev). b) Kerr microcomb as a frequency divider: fully-active approach (1) introduces coupling of the feedback loops, active-passive approach with a fixed pump (2) provides orthogonalization, but limits the flexibility of the operation, and flexible active-passive approach (3) enables the control of any two microcomb teeth, being the most advantageous of the three. c) Flexible stabilization description: The full microcomb stabilization is achieved starting by a "single-pinned" state, where an external reference injection-locks a tooth at $\mu = 0$, followed by a "double-pinned" state, where an active feedback loop locks a second tooth at $\mu = A$. Instead of directly stabilizing the pump ($\mu = P$), the pump acts as an actuator to control $f_r$. This eliminates constraints on the pump frequency, allows flexible locking of any comb tooth, simplifies clock architecture, and maximizes optical division bandwidth. In the single-pinned state the pulse-to-pulse delay $\tau$ fluctuates, whereas in the double-pinned state $\tau$ remains constant.}  
    \label{fig:concept}
\end{figure*}

At the forefront of frequency stability are optical lattice clocks~\cite{yang2025clock} and single-ion clocks~\cite{lindvall202588}, operating at relative frequency stability levels of $10^{-16}$ and $10^{-15}$, respectively. Iodine clocks offer a more compact, albeit less stable, alternative, typically operating at the $10^{-14}$ level. Achieving such levels of precision, commonly characterized by the Allan deviation (Fig.~\ref{fig:concept}[a]), requires sophisticated engineering of both the atomic reference and the optical frequency division based on frequency combs. This complexity generally results in bulky architectures that are challenging to deploy outside the laboratory.

However, advances in photonic integration have enabled substantial progress toward the miniaturization of optical atomic clocks. The development of miniaturized atomic references \cite{kitching2018chip,Isichenko2023,Romaszko2020}, together with integrated Kerr microcombs~\cite{chang2022integrated}, naturally suggests optical-clock architectures suitable for on-chip implementation.  Though not as accurate as their bigger counterparts, the miniaturized atomic references, mostly represented by vapor-cells, can achieve a $10^{-13}$ level of relative frequency stability in a two-photon absorption configuration~\cite{callejo2024short}, putting them potentially on par with the modern bulky hydrogen masers. In particular, silicon nitride (SiN) microring resonators (MRRs), which enable compact soliton Kerr microcomb generation~\cite{herr2016dissipative}, have attracted broad interest owing to their small footprint and applications across metrology, sensing, ranging, quantum technologies, and communications~\cite{sun2023applications,kues2019quantum,diakonov2024broadband,stern2020direct}. Fully stabilized microcombs have also been demonstrated~\cite{del2008full}, positioning them as central components of integrated photonic atomic clocks.

To transfer the stability of an optical reference using a Kerr microcomb, its two degrees of freedom, the carrier-envelope offset frequency ($f_{\mathrm{CEO}}$) and repetition rate ($f_r$), must be controlled. This can be achieved by locking the frequencies of two comb teeth, a method referred to as double-pinning~\cite{li2014electro}. Full optical-frequency division typically relies on $f$-$2f$ interferometry \cite{telle1999carrier}, which requires an octave-spanning comb and remains challenging for integrated platforms \cite{pfeiffer2017octave,song2024octave}. Alternatively, double-pinning enables partial division, with the division factor determined by the spectral separation between the locked teeth \cite{he2024chip,kudelin2024photonic,sun2024integrated}. Conventionally, active feedback loops are used to stabilize both degrees of freedom \cite{newman2019architecture, drake2019terahertz} as shown in Fig.~\ref{fig:concept}[b1], but this approach introduces control complexity and cross-coupling that hinder integration. Injection-locking-based Kerr-induced synchronization (Fig.~\ref{fig:concept}[b2]) has recently emerged as an elegant alternative \cite{moille2023kerr,wildi2023sideband}, enabling passive stabilization of one comb tooth, suppression of comb-line noise, and increased flexibility in frequency selection \cite{moille2025versatile,moille2025all}. Yet, bringing integrated Kerr microcombs into the performance regime of optical lattice and single-ion clocks requires advanced stabilization architectures capable of suppressing residual instability over extended timescales. This raises the question of whether a compact clockwork based on a microcomb (Fig.~\ref{fig:concept}[a]) can rise to this challenge.

Here, we demonstrate that hybrid passive-active stabilization of a Kerr microcomb by injection-locking one comb tooth and actively stabilizing a second, freely selected tooth through pump-frequency actuation, can indeed transfer frequency stability at the level required by state-of-the-art optical clocks. The separation of the passive and active control channels effectively orthogonalizes stabilization of the two microcomb frequency degrees of freedom, while avoiding the requirement to directly lock the pump frequency. This architecture enables flexible double-pinning across the available microcomb bandwidth and is compatible with both partial optical division and, for octave-spanning combs, full $f$-$2f$ self-referencing. Using a continuous five-hour frequency record without post-processed drift removal, we measure a residual relative frequency instability of $1.4\times10^{-16}$ at 1 s, averaging down to $2\times10^{-18}$ at 4000 s on an out-of-loop comb tooth. The residual instability between 1 and 100 s remains unchanged as the injection power is varied over 14 dB, demonstrating robust operation across a broad locking range. The pump actuator exhibits substantially larger frequency excursions than the out-of-loop tooth, showing that the actuator motion does not directly represent the residual motion of the stabilized comb grid. Hybrid locking remains operational for more than five hours, with the measurement duration ultimately limited by the soliton lifetime. We also demonstrate precise control of the pump actuator by means of injection-locking via an hour-long programmable frequency steering. These results bring integrated microcomb stabilization into the residual-noise regime required to preserve the long-term performance of emerging optical lattice and single-ion clock systems.

\begin{figure*}[t]
    \centering
    \includegraphics[width=0.95\linewidth]{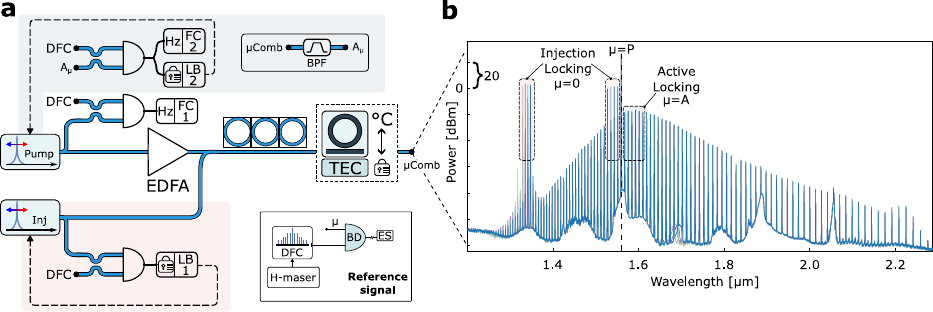}
    \caption{\textbf{Experimental arrangement and Soliton Microcomb Spectrum.} a) Schematic depiction of the setup: The pump laser is amplified by an EDFA and coupled to the SiN MRR for soliton generation. A filtered microcomb tooth $\mu=A$ is stabilized by non-linear flexible active mechanism (see Fig.\ref{fig:concept}[c]). The injection laser is coupled into the cavity to passively lock the $\mu=0$ comb tooth via injection-locking. Both the injection laser and the actively stabilized microcomb tooth are referenced to a hydrogen maser via a tabletop fiber-based frequency comb (DFC). The pump laser beat and the actively-stabilized tooth beat are monitored by frequency counters (FC) for stability characterization. b) Soliton microcomb spectrum generated by MRR: Microcomb mode number $\mu=0$ is assigned to the injection-locked tooth, pump mode number is $\mu=P$, and the actively-stabilized tooth is assigned the mode number $\mu = A$. The light-red and gray highlighted spectral regions indicate the experimentally accessible ranges for passive injection locking and active stabilization, respectively, determined in the present setup by the tuning range of the available CTL lasers and the optical band-pass filter.}.
    \label{fig:schematic}
\end{figure*}

\section*{Results}
\subsection*{Locking flexibility}

We consider a Kerr microresonator-based microcomb whose degrees of freedom, the carrier-envelope offset frequency and the repetition rate, remain unstabilized. The comb structure is described by the equation $f_\mu = f_0 + \mu f_r$, where $f_\mu$ denotes the frequency of the $\mu$-th comb line, offset from a reference frequency $f_0$ by an integer multiple $\mu$ of the repetition rate $f_r$. In the time-domain the comb spectrum corresponds to a train of pulses separated by the delay $\tau$, which is reciprocal to the repetition rate (see Fig.\ref{fig:concept}[c]). Central to this work is the stabilization of two independent degrees of freedom of the comb, which is equivalent to locking any two of its teeth, a process often referred to as full stabilization. To achieve full stabilization, apart from the pump laser used for the comb generation, we utilize an additional laser that we inject into the MRR cavity (Fig.\ref{fig:concept}[a]). We define $f_0$ as the frequency of the injection laser assigning it the mode number $\mu=0$. Prior to injection, the additional laser is stabilized to an optical reference, such as an atomic transition or a high-Q cavity resonance. Then, upon injection, we set the frequency of the injection laser near the target comb line with mode number $\mu=0$ (Fig.\ref{fig:concept}[c]). When the injection laser is tuned sufficiently close to the target comb line and enters the locking range (LR), the target line becomes frequency-locked to the injection source. As a result, the comb tooth passively aligns with the injection laser via the injection-locking effect \cite{adler1946study}, thereby stabilizing one degree of freedom in a regime we refer to as the 'single-pinned' state. Prior to injection-locking, tuning the pump frequency affects both the reference frequency $f_0$ and the repetition rate $f_r$, due to thermo-optic and nonlinear coupling \cite{stone2018thermal}. Once injection-locking is established, the fluctuations in the pump frequency no longer shift the locked comb tooth ($f_0$), and instead solely drive changes in $f_r$. Note, that directly stabilizing the pump frequency would stabilize the microcomb, but with a fixed division factor determined by the frequency offset between the pump and the injection-locked laser. Instead, we use the pump as a mediator to flexibly control and stabilize the frequency of any desired comb tooth, allowing substantially higher division factors. To realize this approach, we leverage the microcomb's inherent frequency relation to fix its second degree of freedom by actively stabilizing a second comb tooth at $\mu = A$, allowing flexible control. During the active-locking stage, a second optical reference is used to create an error signal (ES) with respect to the comb line around $\mu=A$, which is then fed back to the pump's frequency. The pump's frequency, acting as a ‘knob’, stabilizes the tooth with $\mu=A$, thereby fixing the repetition rate and achieving what we term the "double-pinned" state. In the double-pinned state, the stability of the reference sources is transferred to the microcomb’s degrees of freedom. 

\subsection*{Experimental Verification}

\begin{figure*}[h!]
    \centering
    \includegraphics[width=0.9\linewidth]{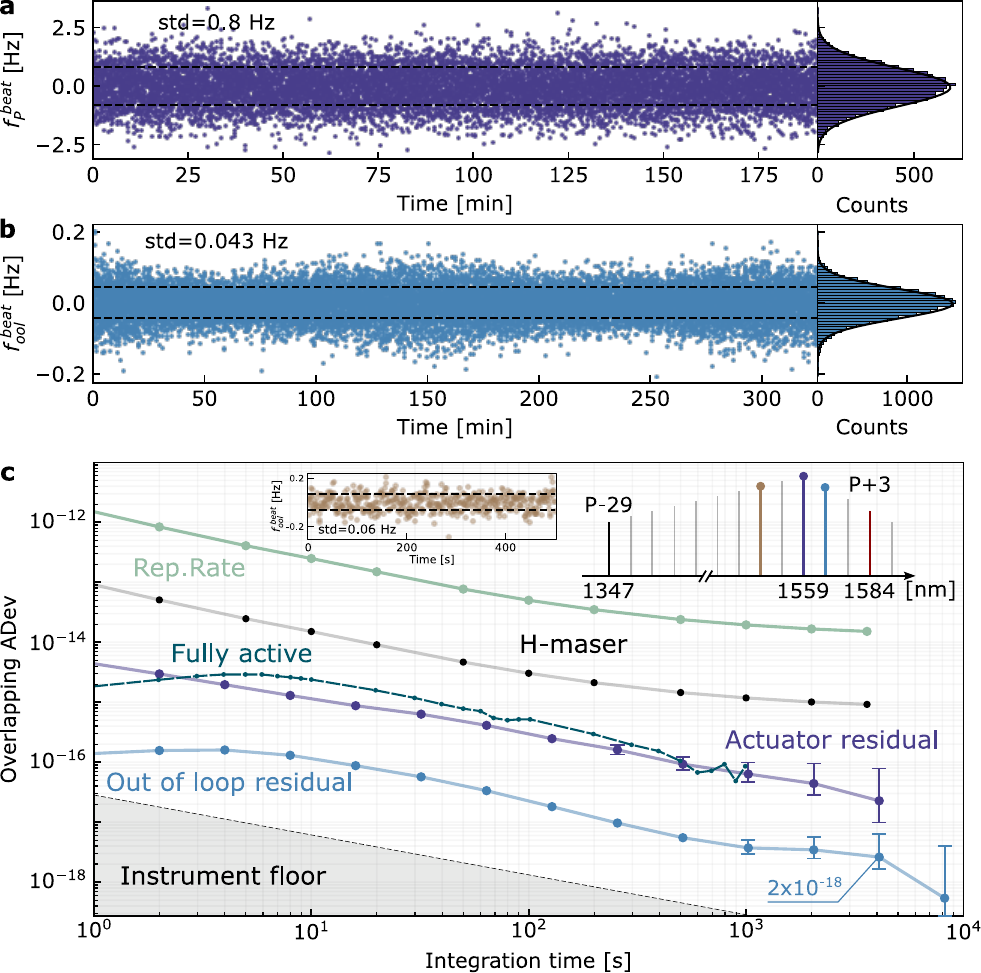}
    \caption{\textbf{Residual stability transfer measurement in flexible hybridly-locked microcomb}: (a-b): Time traces from frequency counter when both teeth with $\mu=P+3$ (active-locking) and $\mu=P-29$ (passive-locking) are tightly stabilized with respect to H-maser referenced DFC. Actuator pump tooth beat [a] and out-of-loop tooth beat [b] with the standard deviations of 0.8 Hz and 0.043 Hz, correspondingly. c) Overlapping Allan Deviation measurements: pump actuator residual  showing $\approx 3 \times 10^{-15}$ and $1 \times 10^{-17}$ at 1 and 2000 s, correspondingly; out-of-loop tooth residual showing $\approx 1.4 \times 10^{-16}$ and $2 \times 10^{-18}$ at 1 and 2000 s, correspondingly. The inset shows the relative positions of actively and passively locked teeth for clarity. For comparison, the residual instability measured in a fully-active setup \cite{drake2019terahertz} is also shown.}
    \label{fig:strong_locking}
\end{figure*}

A simplified experimental setup is shown in Fig.\ref{fig:schematic}[a]. Two tunable lasers were used as the pump and injection sources, respectively. The pump laser, operating at a wavelength of 1559 nm, was amplified by an Erbium-Doped Fiber Amplifier (EDFA) and coupled into the silicon nitride microring resonator (MRR) via a bus waveguide to enable Kerr microcomb generation. Light was coupled into and out of the chip via lensed fibers, while a fiber-based polarization controller (PC) optimized polarization alignment with the MRR bus waveguide mode. The chip temperature was stabilized using a thermo-electric controller (see \cite{diakonov2024broadband} for more details). To stabilize both the pump and injection lasers, we used a commercial table-top difference-frequency comb (DFC) referenced to a hydrogen maser atomic clock. For active locking, we generated a beat note by heterodyning the DFC with one of the microcomb teeth located on the higher‑frequency side of the pump ($\mu=A$). This beat was isolated with an optical band‑pass filter (BPF) and used to generate an error signal (ES), which was then fed into the locking electronics to control the pump laser. Importantly, the pump itself was not directly stabilized; instead, the error signal was used to adjust the pump frequency, compensating for fluctuations of the microcomb tooth at $\mu = A$. We also note that, in principle, stabilization could equally be performed using a tooth on either frequency side of the pump. After establishing the single-pinned regime, we employed injection-locking to stabilize a second microcomb tooth at $\mu = 0$. The injection laser, stabilized to the DFC (and thereby referenced to a hydrogen maser), was tuned into the locking range (LR) around the tooth at $\mu = 0$, at which point it became part of the microcomb. This resulted in a successful transition to the "double-pinned" state. Throughout the experiment, beat notes with the DFC, from both the pump and the actively stabilized tooth at $\mu = A$, were continuously monitored. 

Our ultimate goal was to evaluate the fundamental stability of the active-passive optical atomic clock architecture and thereby elucidate its stability-transfer capabilities. To this end, we sought to isolate the residual frequency instability associated with the two comb teeth used to pin the microcomb. These residual measurements provide a stringent test of the stability-transfer mechanism and reveal its performance limits. To implement this measurement, the pump frequency is compared with a DFC referenced to a hydrogen maser, which provides the stable RF reference required for precise optical-frequency tracking. Importantly, while the maser is essential for establishing this measurement framework, its frequency instability contributes negligibly to the measured pump instability, as shown in the derivation (see Methods). The measurement therefore directly probes the relationship between the residual fluctuations of the pump laser ($\mu = P$), the actively stabilized comb tooth ($\mu = A$), and the injection-locked tooth ($\mu = 0$):

\begin{flalign}
\delta f_P^{beat} = \left\{ \frac{A-P}{A} \delta f_0^{res} + \frac{P}{A} \delta f_A^{res} \right\},
\label{eq}
\end{flalign}

Here, $\delta f_P^{\text{beat}}$ denotes the measured variation in the pump beat, $\delta f_0^{\text{res}}$ represents the residual fluctuation of the injection-locked beat, and $\delta f_A^{\text{res}}$ corresponds to the residual fluctuation of the beat from the actively stabilized tooth. All of these quantities can be directly measured without requiring comparison to an independent ultra-stable reference. We have experimentally verified Eq.(\ref{eq}) by initializing the "double-pinning" state and measuring the ratio between the frequency deviations of the pump and actively-locked tooth. The resulting ratio showed a very close agreement with the analytical expression (see Supplementary material). The "double-pinning" regime with an active stabilization was achieved for the comb teeth up to $P+6$ tooth ($~\approx$1610 nm, see Fig.\ref{fig:schematic}[b]), where we were limited only by the available tunable filters required for the tooth beat extraction. It is important to emphasize that the squared hyperbolic-secant spectral envelope of the soliton can complicate beat-note extraction for active locking, since the comb-tooth power decreases with increasing spectral distance from the pump. This limitation can be mitigated by dispersion engineering, for example by positioning a dispersive wave near the desired locking wavelength to provide a sufficiently strong comb tooth and improve the beat-note signal-to-noise ratio. Alternatively, a weak comb tooth can be optically amplified by injection-locking a low-cost diode laser to that tooth. In contrast, the passive-locking part of our approach does not suffer from the same beat-note signal-to-noise limitation. As shown in Fig.~\ref{fig:schematic}[b], injection-locking was successfully implemented in the $1.34~\mu\text{m}$ spectral region, where the soliton spectral power was lowest.

\begin{figure*}[t]
    \centering
    \includegraphics[width=\linewidth]{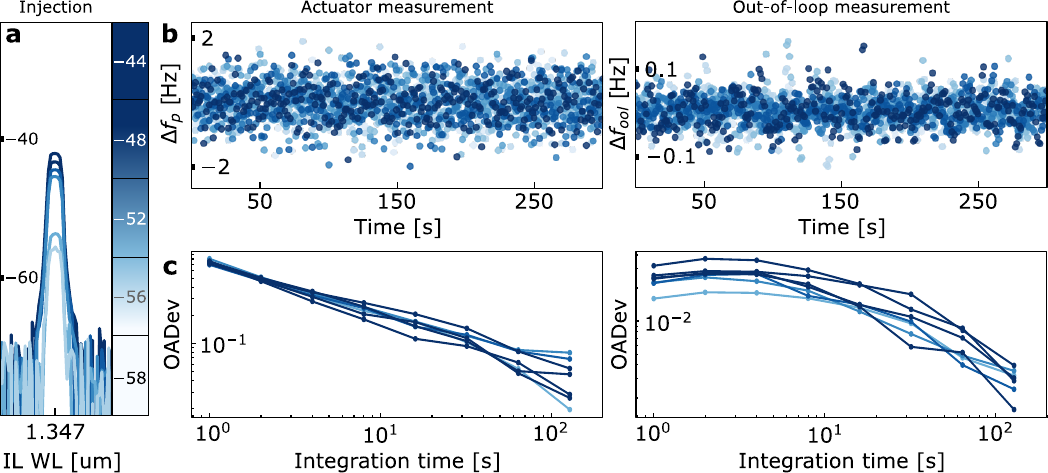}
    \caption{Relative frequency stability as a function of injection power. (a) Injection powers corresponding to the subsequent stability measurements. (b): time traces of the pump actuator and out-of-loop tooth residual frequency deviations, respectively. (c): corresponding OADevs. No significant dependence of the frequency stability on injection power is observed. The apparent instability in the out-of-loop measurement is attributed to environmental variations.}
    \label{fig:power}
\end{figure*}

To identify metrological limits of the stability transfer by means of the hybrid active-passive approach, we focused our attention on its long-term performance. Both the injection laser and active stabilization were implemented with respect to the same DFC, giving the residual stability as a limit of the achievable stability transfer. Active locking was initiated at the 1584 nm tooth and passive injection was applied to the 1347 nm tooth. The pump residual frequency deviation was measured to assess actuator performance; its time trace is shown at Fig.\ref{fig:strong_locking} [a]. Over a period of more than 3 hours, a standard deviation of 0.8 Hz was obtained. Interestingly enough, the out-of-loop measurement of the microcomb tooth (which is neither the pump nor a locked line) showed an even better performance (see Fig.\ref{fig:strong_locking}[b]). A standard deviation of 0.043 Hz was achieved, approaching the measurement noise floor. In order to verify that the chosen out-of-loop tooth was not coincidentally at some 'quiet' optimal point, we performed additional out-of-loop measurement at the tooth around 1543 nm, where the comparable low frequency deviation was achieved as shown in the inset of the Fig.\ref{fig:strong_locking}[c]. This result demonstrates that the injection-locking approach can effectively suppress residual pump fluctuations. To assess the frequency stability, the time traces were converted to the Allan Deviation plots as shown in Fig.\ref{fig:strong_locking} [c], where the uncertainty was calculated assuming the 68\% (one standard deviation) confidence interval. With the value of $1.4\times10^{-16}$ at 1 s integration time and $2\times10^{-18}$ at 4000 s, when normalized with respect to the optical frequency of the out-of-loop tooth (1567 nm), our experiment shows the best residual performance demonstrated so far in the microcomb-based stability transfer. The operational regime can be easily initiated and maintained, while the only limitation is the soliton existence time. We attribute the breathing pattern of the out-of-loop measurement, which translates to a non-$1/\sqrt{\tau}$ trend in the Allan deviation, to the fiber noise induced by the unshielded environmental temperature fluctuations. In case of the pump actuator measurement the standard deviation is much higher, which can mask the noise present in the out-of-loop case, and the OADev achieved (normlized with respect to the 1559 nm pump frequency) is $\approx3\times10^{-15}$ and $\approx1\times10^{-17}$ at 1 and 4000 s, correspondingly.

As injection-locking power defines the locking-range bandwidth, it is crucial to understand its effect on the microcomb stability transfer. In our experiment, we performed a series of 5-minute frequency deviation measurements, while varying the injection power in the range of 14 dB. In line with the multi-hour measurement, we characterized the pump actuator in the hybridly-locked state, as well as the out-of-loop microcomb tooth. The results are shown in Fig.\ref{fig:power}. Importantly, it is worth emphasizing that the proposed approach can operate with the injection-locking power varying from 500 $\mu W$ (corresponding to the output power of -44 dBm as shown in Fig.\ref{fig:power} [a]) down to 20 $\mu W$ (-58 dBm output power), demonstrating reliable control using very modest injection power. In the case of the pump actuator measurement (Fig.\ref{fig:power}[b]), no significant dependence in the whole range of injection powers was found. As for the case of the out-of-loop measurement, a slight dependence in frequency deviation behavior was revealed, where a few different measurements did not show a consistent trend. As in the case of the long-term out-of-loop measurement, we attribute this inconsistency to the ambient temperature fluctuations. 

\begin{figure}[t]
     \centering
    \includegraphics[width=\linewidth]{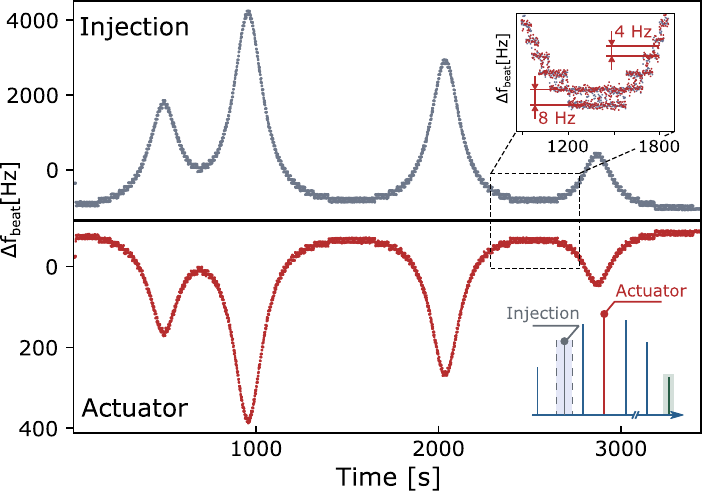}
    \caption{\textbf{Demonstration of precise pump-actuator control via hybrid locking.} The injection-laser frequency (top) is programmed to follow the profile of the natural rubidium-vapor spectrum in discrete steps of several tens of hertz. The pump actuator (bottom) follows these programmed changes with hertz-level precision over a measurement duration of nearly one hour. The inset highlights the discrete nature of the frequency control, where individual steps are clearly resolved. For ease of comparison, the actuator trace is inverted in the inset.}
    \label{fig:rubidium}
\end{figure}

To showcase the arbitrary frequency steering and high-resolution control enabled by our hybrid approach, we drove the injection laser along a complex frequency trajectory. To connect this demonstration directly to optical metrology, the trajectory was chosen to reproduce the profile of the natural rubidium-vapor spectrum, while the corresponding response of the pump actuator was measured simultaneously, as shown in Fig.~\ref{fig:rubidium}. During a nearly one-hour measurement, we controlled the injection laser (top panel of Fig.\ref{fig:rubidium}) with a precision of tens of Hz, which was translated to the Hz precision of the pump response (bottom panel of Fig.\ref{fig:rubidium}). We emphasize that this measurement simply mimics the absorption spectrum, rather than demonstrates the real experimental atomic spectroscopy. The inset of the figure zooms in into the small time window of the measurement to show the discrete nature of the frequency control. For the comparison purposes inside the zoom-in part the actuator plot is reversed to be matched with the injection one.

\section*{Discussion}

In this work, we have demonstrated that a hybrid active-passive stabilization approach can transfer optical frequency stability to a microcomb while maintaining long-term operation over several hours. The stabilized comb reaches a residual relative frequency instability of $2\times10^{-18}$ at an averaging time of $4000\text{ s}$, demonstrating that the locking scheme is not limited to short-term suppression of frequency fluctuations, but preserves high stability over timescales relevant to precision metrology. This residual instability is well below the stability requirements of many compact optical frequency references and demonstrates that the hybrid locking architecture itself need not limit the performance of advanced optical metrological systems.

A notable feature of the measurements is that the out-of-loop comb teeth exhibit lower frequency noise and improved long-term stability compared with the directly measured pump. This behavior is reproduced in the measurements of two individual comb teeth, indicating that it is not associated with a single most optimal measurement point. The pump frequency is measured directly after the ECDL and before coupling into the microresonator, whereas the out-of-loop teeth are generated only after the pump interacts with the microresonator and undergoes nonlinear comb generation. Therefore we hypothesize that the two measurements probe different responses of the stabilized system, and fluctuations observed on the pump are not necessarily transferred one-to-one to the generated comb modes. The resulting frequency-noise spectra can consequently differ, even though the optical frequencies remain linked through the comb dynamics. More generally, non-trivial noise transfer in injection-locked microcombs has also been reported previously~\cite{moille2025all}.

Beyond the achieved stability, an important advantage of the hybrid scheme is its spectral flexibility. By decoupling the stabilization process from the pump laser, a selected microcomb tooth can be referenced directly to an optical transition without requiring the reference frequency to lie close to the pump. This potentially opens the full usable bandwidth of the comb for stabilization and allows the system to interface with a substantially broader range of optical frequency references.

This flexibility is particularly useful for optical frequency division. In partial-division schemes, simultaneously matching the pump frequency and a secondary comb tooth to two independent optical references can otherwise impose restrictive spectral constraints. The present approach largely relaxes this requirement by providing an additional degree of freedom in selecting the stabilized comb tooth. The same principle could also complement low-phase-noise microwave-generation techniques based on Kerr-induced synchronization, such as the optical division scheme demonstrated by Sun et al.~\cite{sun2024kerr}, by extending the optical bandwidth over which external references can be interfaced with the microcomb.

The demonstrated approach is therefore compatible with a broad range of optical frequency references, spanning compact systems such as micromachined vapor cells with sub-$10^{-13}$ stability at $1\text{ s}$~\cite{kitching2018chip} to substantially more stable references based on single alkaline-earth ions. Importantly, long-term operation was maintained without specialized suppression of environmental perturbations, while stable injection control was achieved with only $20~\mu\text{W}$ of optical power and without measurable degradation of the residual stability. Taken together, the combination of several-hour operation, residual instability reaching the low-$10^{-18}$ regime, low required injection power, and broad spectral flexibility demonstrates the suitability of hybrid active-passive stabilization for compact optical frequency references and optical frequency-division systems. These characteristics are particularly relevant for the development of miniaturized optical atomic clocks and other precision metrological systems intended for operation outside specialized laboratory environments.

\section*{Acknowledgments}
The authors thank Gregory Moille for insightful feedback.

\clearpage
\newpage
\bibliographystyle{abbrv}
\bibliography{apssamp}

@article{ludlow2015optical,
  title={Optical atomic clocks},
  author={Ludlow, Andrew D and Boyd, Martin M and Ye, Jun and Peik, Ekkehard and Schmidt, Piet O},
  journal={Reviews of Modern Physics},
  volume={87},
  number={2},
  pages={637--701},
  year={2015},
  publisher={APS}
}

@article{mcgrew2018atomic,
  title={Atomic clock performance enabling geodesy below the centimetre level},
  author={McGrew, WF and Zhang, X and Fasano, RJ and Sch{\"a}ffer, SA and Beloy, K and Nicolodi, D and Brown, RC and Hinkley, N and Milani, G and Schioppo, M and others},
  journal={Nature},
  volume={564},
  number={7734},
  pages={87--90},
  year={2018},
  publisher={Nature Publishing Group UK London}
}

@article{blewitt2009gps,
  title={GPS for real-time earthquake source determination and tsunami warning systems},
  author={Blewitt, Geoffrey and Hammond, William C and Kreemer, Corn{\'e} and Plag, Hans-Peter and Stein, Seth and Okal, Emile},
  journal={Journal of geodesy},
  volume={83},
  pages={335--343},
  year={2009},
  publisher={Springer}
}

@article{safronova2019search,
  title={The search for variation of fundamental constants with clocks},
  author={Safronova, Marianna S},
  journal={Annalen der Physik},
  volume={531},
  number={5},
  pages={1800364},
  year={2019},
  publisher={Wiley Online Library}
}

@article{herr2016dissipative,
  title={Dissipative Kerr solitons in optical microresonators},
  author={Herr, Tobias and Gorodetsky, Michael L and Kippenberg, Tobias J},
  journal={Nonlinear optical cavity dynamics: from microresonators to fiber lasers},
  pages={129--162},
  year={2016},
  publisher={Wiley Online Library}
}

@article{sun2023applications,
  title={Applications of optical microcombs},
  author={Sun, Yang and Wu, Jiayang and Tan, Mengxi and Xu, Xingyuan and Li, Yang and Morandotti, Roberto and Mitchell, Arnan and Moss, David J},
  journal={Advances in Optics and Photonics},
  volume={15},
  number={1},
  pages={86--175},
  year={2023},
  publisher={Optica Publishing Group}
}

@article{kues2019quantum,
  title={Quantum optical microcombs},
  author={Kues, Michael and Reimer, Christian and Lukens, Joseph M and Munro, William J and Weiner, Andrew M and Moss, David J and Morandotti, Roberto},
  journal={Nature Photonics},
  volume={13},
  number={3},
  pages={170--179},
  year={2019},
  publisher={Nature Publishing Group UK London}
}

@article{diakonov2024broadband,
  title={Broadband cavity-enhanced Kerr comb spectroscopy on chip},
  author={Diakonov, Andrei and Khrizman, Konstantin and Zano, Eliran and Stern, Liron},
  journal={npj Nanophotonics},
  volume={1},
  number={1},
  pages={1--9},
  year={2024},
  publisher={Nature Publishing Group}
}

@article{stern2020direct,
  title={Direct Kerr frequency comb atomic spectroscopy and stabilization},
  author={Stern, Liron and Stone, Jordan R and Kang, Songbai and Cole, Daniel C and Suh, Myoung-Gyun and Fredrick, Connor and Newman, Zachary and Vahala, Kerry and Kitching, John and Diddams, Scott A and others},
  journal={Science advances},
  volume={6},
  number={9},
  pages={eaax6230},
  year={2020},
  publisher={American Association for the Advancement of Science}
}

@article{telle1999carrier,
  title={Carrier-envelope offset phase control: A novel concept for absolute optical frequency measurement and ultrashort pulse generation},
  author={Telle, Harald R and Steinmeyer, Guenter and Dunlop, Amy E and Stenger, Joern and Sutter, Dirk H and Keller, Ursula},
  journal={Applied Physics B},
  volume={69},
  pages={327--332},
  year={1999},
  publisher={Springer}
}

@article{pfeiffer2017octave,
  title={Octave-spanning dissipative Kerr soliton frequency combs in Si3N4 microresonators},
  author={Pfeiffer, Martin HP and Herkommer, Clemens and Liu, Junqiu and Guo, Hairun and Karpov, Maxim and Lucas, Erwan and Zervas, Michael and Kippenberg, Tobias J},
  journal={Optica},
  volume={4},
  number={7},
  pages={684--691},
  year={2017},
  publisher={Optical Society of America}
}

@article{song2024octave,
  title={Octave-spanning Kerr soliton frequency combs in dispersion-and dissipation-engineered lithium niobate microresonators},
  author={Song, Yunxiang and Hu, Yaowen and Zhu, Xinrui and Yang, Kiyoul and Lon{\v{c}}ar, Marko},
  journal={Light: Science \& Applications},
  volume={13},
  number={1},
  pages={225},
  year={2024},
  publisher={Nature Publishing Group UK London}
}

@article{he2024chip,
  title={Chip-scale high-performance photonic microwave oscillator},
  author={He, Yang and Cheng, Long and Wang, Heming and Zhang, Yu and Meade, Roy and Vahala, Kerry and Zhang, Mian and Li, Jiang},
  journal={Science Advances},
  volume={10},
  number={33},
  pages={eado9570},
  year={2024},
  publisher={American Association for the Advancement of Science}
}

@article{kudelin2024photonic,
  title={Photonic chip-based low-noise microwave oscillator},
  author={Kudelin, Igor and Groman, William and Ji, Qing-Xin and Guo, Joel and Kelleher, Megan L and Lee, Dahyeon and Nakamura, Takuma and McLemore, Charles A and Shirmohammadi, Pedram and Hanifi, Samin and others},
  journal={Nature},
  volume={627},
  number={8004},
  pages={534--539},
  year={2024},
  publisher={Nature Publishing Group UK London}
}

@article{sun2024integrated,
  title={Integrated optical frequency division for microwave and mmWave generation},
  author={Sun, Shuman and Wang, Beichen and Liu, Kaikai and Harrington, Mark W and Tabatabaei, Fatemehsadat and Liu, Ruxuan and Wang, Jiawei and Hanifi, Samin and Morgan, Jesse S and Jahanbozorgi, Mandana and others},
  journal={Nature},
  volume={627},
  number={8004},
  pages={540--545},
  year={2024},
  publisher={Nature Publishing Group UK London}
}

@article{drake2019terahertz,
  title={Terahertz-rate Kerr-microresonator optical clockwork},
  author={Drake, Tara E and Briles, Travis C and Stone, Jordan R and Spencer, Daryl T and Carlson, David R and Hickstein, Daniel D and Li, Qing and Westly, Daron and Srinivasan, Kartik and Diddams, Scott A and others},
  journal={Physical Review X},
  volume={9},
  number={3},
  pages={031023},
  year={2019},
  publisher={APS}
}

@article{moille2023kerr,
  title={Kerr-induced synchronization of a cavity soliton to an optical reference},
  author={Moille, Gr{\'e}gory and Stone, Jordan and Chojnacky, Michal and Shrestha, Rahul and Javid, Usman A and Menyuk, Curtis and Srinivasan, Kartik},
  journal={Nature},
  volume={624},
  number={7991},
  pages={267--274},
  year={2023},
  publisher={Nature Publishing Group UK London}
}

@article{stone2018thermal,
  title={Thermal and nonlinear dissipative-soliton dynamics in Kerr-microresonator frequency combs},
  author={Stone, Jordan R and Briles, Travis C and Drake, Tara E and Spencer, Daryl T and Carlson, David R and Diddams, Scott A and Papp, Scott B},
  journal={Physical review letters},
  volume={121},
  number={6},
  pages={063902},
  year={2018},
  publisher={APS}
}

@article{moille2025versatile,
  title={Versatile optical frequency division with Kerr-induced synchronization at tunable microcomb synthetic dispersive waves},
  author={Moille, Gr{\'e}gory and Shandilya, Pradyoth and Niang, Alioune and Menyuk, Curtis and Carter, Gary and Srinivasan, Kartik},
  journal={Nature Photonics},
  volume={19},
  number={1},
  pages={36--43},
  year={2025},
  publisher={Nature Publishing Group UK London}
}

@article{sun2024kerr,
  title={Kerr optical frequency division with integrated photonics for stable microwave and mmWave generation},
  author={Sun, Shuman and Harrington, Mark W and Tabatabaei, Fatemehsadat and Hanifi, Samin and Liu, Kaikai and Wang, Jiawei and Wang, Beichen and Yang, Zijiao and Liu, Ruxuan and Morgan, Jesse S and others},
  journal={arXiv preprint arXiv:2402.11772},
  year={2024}
}

@article{adler1946study,
  title={A study of locking phenomena in oscillators},
  author={Adler, Robert},
  journal={Proceedings of the IRE},
  volume={34},
  number={6},
  pages={351--357},
  year={1946},
  publisher={IEEE}
}

@article{yang2025clock,
  title={Clock precision beyond the Standard Quantum Limit at $10^{-18}$ level},
  author={Yang, YA and Miklos, Maya and Tso, Yee Ming and Kraus, Stella and Hur, Joonseok and Ye, Jun},
  journal={arXiv preprint arXiv:2505.04538},
  year={2025}
}

@article{del2008full,
  title={Full stabilization of a microresonator-based optical frequency comb},
  author={Del’Haye, Pascal and Arcizet, Olivier and Schliesser, Albert and Holzwarth, Ronald and Kippenberg, Tobias J},
  journal={Physical Review Letters},
  volume={101},
  number={5},
  pages={053903},
  year={2008},
  publisher={APS}
}

@article{newman2019architecture,
  title={Architecture for the photonic integration of an optical atomic clock},
  author={Newman, Zachary L and Maurice, Vincent and Drake, Tara and Stone, Jordan R and Briles, Travis C and Spencer, Daryl T and Fredrick, Connor and Li, Qing and Westly, Daron and Ilic, Bojan R and others},
  journal={Optica},
  volume={6},
  number={5},
  pages={680--685},
  year={2019},
  publisher={Optical Society of America}
}

@article{wildi2023sideband,
  title={Sideband injection locking in microresonator frequency combs},
  author={Wildi, Thibault and Ulanov, Alexander and Englebert, Nicolas and Voumard, Thibault and Herr, Tobias},
  journal={APL photonics},
  volume={8},
  number={12},
  year={2023},
  publisher={AIP Publishing}
}

@article{kitching2018chip,
  title={Chip-scale atomic devices},
  author={Kitching, John},
  journal={Applied Physics Reviews},
  volume={5},
  number={3},
  year={2018},
  publisher={AIP Publishing}
}

@article{callejo2024short,
  title={Short-term stability of a microcell optical reference based on the Rb atom two-photon transition at 778 nm},
  author={Callejo, Martin and Mursa, Andrei and Vicarini, R{\'e}my and Klinger, Emmanuel and Tanguy, Quentin and Millo, Jacques and Passilly, Nicolas and Boudot, Rodolphe},
  journal={Journal of the Optical Society of America B},
  volume={42},
  number={1},
  pages={151--159},
  year={2024},
  publisher={Optica Publishing Group}
}

@article{guo2017universal,
  title={Universal dynamics and deterministic switching of dissipative Kerr solitons in optical microresonators},
  author={Guo, Hairun and Karpov, Maxim and Lucas, Erwan and Kordts, Arne and Pfeiffer, Martin HP and Brasch, Victor and Lihachev, Grigory and Lobanov, Valery E and Gorodetsky, Michael L and Kippenberg, Tobias J},
  journal={Nature Physics},
  volume={13},
  number={1},
  pages={94--102},
  year={2017},
  publisher={Nature Publishing Group UK London}
}

@article{carmon2004dynamical,
  title={Dynamical thermal behavior and thermal self-stability of microcavities},
  author={Carmon, Tal and Yang, Lan and Vahala, Kerry J},
  journal={Optics express},
  volume={12},
  number={20},
  pages={4742--4750},
  year={2004},
  publisher={Optical Society of America}
}

@article{Isichenko2023,
author={Isichenko, Andrei
and Chauhan, Nitesh
and Bose, Debapam
and Wang, Jiawei
and Kunz, Paul D.
and Blumenthal, Daniel J.},
title={Photonic integrated beam delivery for a rubidium 3D magneto-optical trap},
journal={Nature Communications},
year={2023},
month={May},
day={29},
volume={14},
number={1},
pages={3080},
issn={2041-1723},
doi={10.1038/s41467-023-38818-6},
url={https://doi.org/10.1038/s41467-023-38818-6}
}

@article{Romaszko2020,
author={Romaszko, Zak David
and Hong, Seokjun
and Siegele, Martin
and Puddy, Reuben Kahan
and Lebrun-Gallagher, Foni Rapha{\"e}l
and Weidt, Sebastian
and Hensinger, Winfried Karl},
title={Engineering of microfabricated ion traps and integration of advanced on-chip features},
journal={Nature Reviews Physics},
year={2020},
month={Jun},
day={01},
volume={2},
number={6},
pages={285-299},
issn={2522-5820},
doi={10.1038/s42254-020-0182-8},
url={https://doi.org/10.1038/s42254-020-0182-8}
}

@article{moille2025all,
  title={All-optical noise quenching of an integrated frequency comb},
  author={Moille, Gr{\'e}gory and Shandilya, Pradyoth and Stone, Jordan and Menyuk, Curtis and Srinivasan, Kartik},
  journal={Optica},
  volume={12},
  number={7},
  pages={1020--1030},
  year={2025},
  publisher={Optica Publishing Group}
}

@article{lindvall202588,
  title={88 Sr+ optical clock with 7.9$\times$ 10- 19 systematic uncertainty and measurement of its absolute frequency with 9.8$\times$ 10- 17 uncertainty},
  author={Lindvall, Thomas and Fordell, Thomas and Hanhij{\"a}rvi, KJ and Dole{\v{z}}al, M and Rahm, J and Weyers, S and Wallin, AE},
  journal={Physical Review Applied},
  volume={24},
  number={4},
  pages={044082},
  year={2025},
  publisher={APS}
}

@article{chang2022integrated,
  title={Integrated optical frequency comb technologies},
  author={Chang, Lin and Liu, Songtao and Bowers, John E},
  journal={Nature Photonics},
  volume={16},
  number={2},
  pages={95--108},
  year={2022},
  publisher={Nature Publishing Group UK London}
}

@article{li2014electro,
  title={Electro-optical frequency division and stable microwave synthesis},
  author={Li, Jiang and Yi, Xu and Lee, Hansuek and Diddams, Scott A and Vahala, Kerry J},
  journal={Science},
  volume={345},
  number={6194},
  pages={309--313},
  year={2014},
  publisher={American Association for the Advancement of Science}
}

\newpage
\onecolumn
\section*{Double-pinning mode of operation}

If we successfully achieve the double-pinning state while maintaining a negligible contribution from one lock relative to the other, then, according to Eq.(\ref{eq:correllation}), the relation between the pump beat variation $\delta f_P^{beat}$ and either $\delta f_0^{res}$ or $\delta f_A^{res}$ should follow the specific ratio defined by the integers $A$ and $P$. Following that logic, we applied weak active-locking with an average deviation of a few kHz for $\delta f_A^{res}$, while simultaneously stabilizing the injection laser $\delta f_0^{res}$ to the mHz level, such that the pump beat is expected to follow the ratio $A/P$ according to Eq.(\ref{eq:correllation}). We performed a series of weakly locked measurements during which the first tooth to the right of the pump was actively locked, such that A = P + 1, which results in the slope of $(P+1)/P$.\\

\begin{flalign}
    \delta f_P^{beat} = \left\{ \frac{A-P}{A} \delta f_0^{res} + \frac{P}{A} \delta f_A^{res} \right\},
    \label{eq:correllation}
\end{flalign} 

\begin{figure*}[htp]
    \centering
    \includegraphics[width=0.9\linewidth]{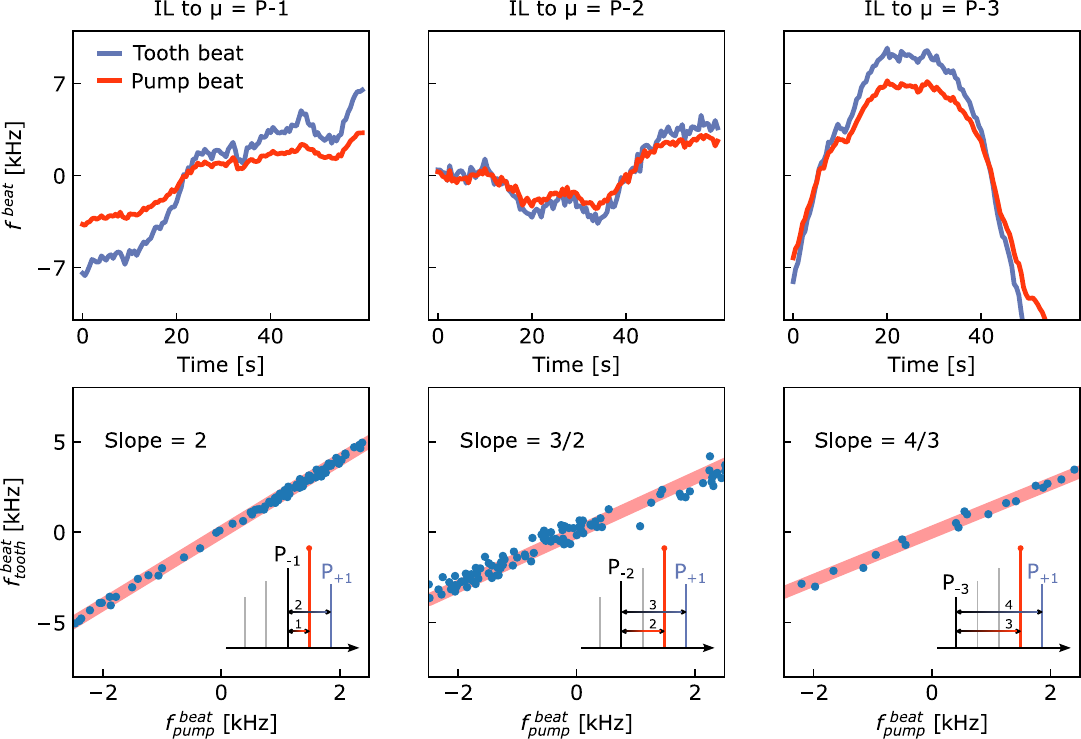}
    \caption{Double-pinning verification by weak-locking measurements: Top row - simultaneous frequency counter measurement of the pump beat (red) and actively-locked tooth with $\mu=P+1$ beat (blue). Bottom row - frequency ratio of the two measurements with the slope corresponding to Eq.\ref{eq:correllation}. Columns from left to right - passive locking of the tooth with $\mu=[P-1, P-2, P-3]$ correspondingly. Each bottom-row inset illustrates the relative positions of the actively and passively locked teeth.}
    \label{fig:weak_locking}
\end{figure*}

The upper panel of Fig.~\ref{fig:weak_locking} shows synchronized measurements of the pump beat and the corresponding actively locked comb-tooth beat, acquired using dead-time-free frequency counters. Their ratio is demonstrated at the bottom part of Fig.\ref{fig:weak_locking} and it clearly obeys  Eq.(\ref{eq:correllation}). For clarity, the insets in Fig.\ref{fig:weak_locking} show the distances between the teeth defining the resulting ratio. Thus, we successfully verified that the system reaches the double-pinning state and can be adjusted to the frequency of the reference source. The measurement fluctuations were primarily caused by uncompensated ambient-temperature variations affecting the active lock. Consequently, the number of data points available for characterizing each ratio varied.

\section*{Stability equation derivation}
We start the derivation from the basic Kerr comb (\ref{eq:Kerr}) and DFC (\ref{eq:DFC}) equations:

\begin{flalign}
f_{\mu} = f_0 + \mu f_r^{Kerr}, \label{eq:Kerr} \\ 
f_{N_{\mu}} = N_{\mu} f_r^{DFC}, \label{eq:DFC}
\end{flalign}

\noindent where $f_0$ denotes injection-locked frequency, $\mu,N_{\mu}$ - longitudinal mode number of Kerr comb and DFC mode number closest to the Kerr mode with number $\mu$, correspondingly, $f_r^{Kerr}, f_r^{DFC}$ - repetition rates of Kerr comb and DFC, correspondingly. Now, we denote Kerr comb mode numbers of actively-locked and pump teeth as $A$ and $P$ getting the equations for $\mu=A$ and $\mu=P$ with their corresponding frequency deviations denoted by $\delta$:

\begin{flalign}
    f_A = f_0 + A f_r^{Kerr} \implies \delta f_A = \delta f_0 + A \delta f_r^{Kerr}, \label{eq:f_A}\\
    f_P = f_0 + P f_r^{Kerr} \implies \delta f_P = \delta f_0 + P \delta f_r^{Kerr} \label{eq:f_P}.
\end{flalign}

To get the dependence of the pump beat frequency deviation on actively-locked tooth beat frequency deviation we start by recasting $f_P$ in terms of $f_0$ using Eq.(\ref{eq:f_A}) and Eq.(\ref{eq:f_P}), which reads as:

\begin{flalign}
    f_P = f_0 \left(\frac{A-P}{A} \right) + \frac{P}{A}f_A. \label{eq:f_0_f_A}
\end{flalign}

Now, we define Kerr comb frequencies with respect to the DFC comb lines:

\begin{flalign}
    f_0 = N_0 f_r^{DFC} + \Delta f_0, \\
    f_A = N_A f_r^{DFC} + \Delta f_A, \\
    f_P = N_P f_r^{DFC} + \Delta f_P,
    \label{eq:f_DFC}
\end{flalign}

\noindent where $N_0, N_A$ and $N_P$ are the DFC mode numbers (integer), and $\Delta f_\mu$ is the offset between the Kerr comb mode and DFC mode including the non-idealities of the locking system and residual noise.

The time variation of $f_0, f_A$ and $f_P$ read as follows:

\begin{flalign}
    \delta f_0 = N_0 \delta f_r^{DFC} + \delta f_0^{res}, \\
    \delta f_A = N_A \delta f_r^{DFC} + \delta f_A^{res}, \\
    \delta f_P = N_P \delta f_r^{DFC} + \delta f_P^{res}, \label{del_f_P}
\end{flalign}

\noindent where the derivative of the constant part of the offset $\Delta f_\mu$ goes to zero leaving only the residual contribution. 

From Eq.\ref{eq:f_0_f_A} and Eq.\ref{del_f_P} we can describe the variation of the pump frequency $f_P$ as follows:
\begin{flalign}
    \delta f_P = \delta f_r^{DFC} \left [N_0\left(1 -  \frac{P}{A} \right) + \frac{P}{A}N_A - N_P \right] + \notag \\
    + \delta f_0^{res} \left( 1 - \frac{P}{A}\right) + \delta f_A^{res} \left( \frac{P}{A} \right).
    \label{eq:delta_fp}
\end{flalign}

To understand the contribution of the right-hand side first term of the Eq.{\ref{eq:delta_fp}} we transform it as follows:

\begin{flalign}
    \delta f_r^{DFC} \left [N_0\left(1 -  \frac{P}{A} \right) + \frac{P}{A}N_A - N_P \right] = \frac{\delta f_r^{DFC}}{f_r^{DFC}} [(P/A)(N_A - N_0)f_r^{DFC} -(N_P - N_0)f_r^{DFC}].
    \label{eq:dfc_contr}
\end{flalign}

Now, $(N_\mu - N_0)f_r^{DFC}$ can be represented in terms of Kerr repetition rate $f_r^{Kerr}$ as $\mu f_r^{Kerr} + \epsilon f_r^{DFC}$, where $\epsilon$ denotes the offset between the Kerr comb line with the number $\mu$ and the closest DFC line in the units of $f_r^{DFC}$ such that $|\epsilon|<1$, thus Eq.(\ref{eq:dfc_contr}) can be recast as:

\begin{flalign}
    \frac{\delta f_r^{DFC}}{f_r^{DFC}} [\frac{P}{A}Af_r^{Kerr} + \frac{P}{A}\epsilon_1 f_r^{DFC} - (Pf_r^{Kerr} + \epsilon_2 f_r^{DFC})] = \delta f_r^{DFC} [\frac{P}{A}\epsilon_1  - \epsilon_2].
    \label{eq:dfc_contr_negl}
\end{flalign}

Getting back to the original Eq.(\ref{eq:delta_fp}) and using Eq.(\ref{eq:dfc_contr_negl} as its RHS we obtain:

\begin{flalign}
    \delta f_P = \delta f_r^{DFC} (\frac{P}{A}\epsilon_1  - \epsilon_2) + \delta f_0^{res} \left( 1 - \frac{P}{A}\right) + \delta f_A^{res} \left( \frac{P}{A} \right).
\end{flalign}

Referencing of the DFC repetition rate to the H-maser results into $\delta f_r^{DFC} \approx 10^{-5}$ Hz, which is negligible with respect to the locking contribution, this brings us to the final equation:

\begin{flalign}
    \delta f_P = \delta f_0^{res} \left( 1 - \frac{P}{A}\right) + \delta f_A^{res} \left( \frac{P}{A} \right).
\end{flalign}

\section*{Chip Design}
Our silicon nitride chip, fabricated by Ligentec, was designed using an 800-nm-thick SiN layer embedded in optical-quality silicon oxide. The chip primarily consists of microring resonators with a radius of \(22.5~\mu\mathrm{m}\) and a waveguide width of approximately \(1500~\mathrm{nm}\), yielding a free spectral range of \(1.06~\mathrm{THz}\) near \(1559~\mathrm{nm}\) for THz-rate frequency-comb generation. A straight bus waveguide with a width of \(1000~\mathrm{nm}\) was used to couple light into the microring. Finite-element-method simulations (FEM, Lumerical) were used to optimize the microring waveguide geometry and obtain a dispersion profile suitable for soliton generation. Finite-difference time-domain simulations (FDTD, Lumerical) were subsequently used to optimize the bus-waveguide geometry and coupling gap for critical coupling, resulting in a gap of \(700~\mathrm{nm}\). Inverse tapers were implemented at the chip facets to improve mode matching to the lensed fibers.

\section*{Experimental Setup}
Two tunable ECDL lasers (Toptica CTL 1550, Toptica CTL 1320) were used for pump control ($\mu = P$) and injection-locking ($\mu=0$), correspondingly. Pump laser with the output power of 1 mW was amplified by Erbium-Doped Fiber Amplifier (EDFA) with the output power of 250 mW entering the chip. Polarization controller followed by a lens fiber was used to couple the amplified light to the chip for the mode-matching purposes, another lens fiber was used for the signal extraction out of the chip. MRR temperature was stabilized to 0.1 $^{\circ}C$ by Thermo-Electric Controller (TEC) managing the Peltier element connected to the rigid copper stage. Pump laser was split by 99/1 fiber coupler, where 99\% of the pump were directed to the EDFA for comb generation and 1\% was used to create feedback for the active control of the arbitrary tooth with $\mu=A$. Difference-Frequency Comb (DFC Toptica Core+) referenced to the H-maser was used as a frequency reference playing the role of the atomic optical transitions. After the successful Kerr comb generation the comb tooth with $\mu=A$ was filtered out by means of fiber Band-Pass filter (BPF) to create a beat with the pump control. This beat was supplied to Toptica Lock-box (PFD Pro + FALC Pro) for Error Signal (ES) generation and subsequent stabilization of the microcomb tooth by means of a pump control. Another Toptica Lock-box was used to stabilize the injection-laser to another DFC tooth. Then 0.5 mW of the injection laser was inserted into the chip via the fiber coupler connected after the EDFA. Offset control of the injection-laser frequency provided by the Lock-box was used to establish injection-locking. The beat between the pump laser and the DFC monitored by Electric Spectrum Analyzer (ESA) was used to identify the injection-locking state by shifting the frequency of the injection-laser until the observation of the quasi injection-locking (sidebands around the beat controlled by the offset between the injection-laser and the comb line \cite{adler1946study}) and subsequent Kerr comb tooth control. Two synchronized gap-free frequency counters (Keysight 53230A) were used to monitor actively-locked beat and the pump beat for the correlation measurements. 

\section*{Soliton initiation}
Our tunable ECDL was slowly tuned from the blue-shifted side of the MRR resonance (1567 nm) towards the red-shifted side to trigger the nonlinear threshold and cascaded four-wave mixing. MRR resonance transmission was monitored on the oscilloscope to obtain the "thermal triangle"\cite{carmon2004dynamical} resonance response, polarization controller was used to optimize the mode-coupling until the soliton step can be observed\cite{herr2016dissipative}. Further, red-shifting resulted in multi-soliton state generation. In order to obtain a single-soliton state we used a backward-tuning technique described in Ref. \cite{guo2017universal} slowly getting back to the blue-side reaching a smooth hyperbolic secant spectrum envelope monitored by OSA.

\end{document}